\documentclass[12pt]{article}
\usepackage{times}
\usepackage{geometry}
\usepackage[utf8]{inputenc}
\usepackage{enumitem,amssymb}
\usepackage{ragged2e}
\newlist{thematic}{itemize}{8}
\setlist[thematic]{label=$\square$}
\usepackage{pifont}
\usepackage{color,slashed}
\usepackage{xcolor}

\usepackage{cite}
\usepackage{endnotes}
\usepackage{manyfoot}
\DeclareNewFootnote{A}[arabic]
\DeclareNewFootnote{B}[alph]

\let\footnote=\endnote

\begin{document}
\newcommand{\Amherst}{University of Massachusetts, Amherst, MA 01003 USA}
\newcommand{\ANLHEP}{HEP Division, Argonne National Laboratory, Lemont, IL 60439, USA}
\newcommand{\APC}{Laboratoire Astroparticule et Cosmologie (APC), CNRS/IN2P3, Universit\'e Paris Diderot, 10, rue Alice Domon et Léonie Duquet, 75205 Paris Cedex 13, France}
\newcommand{\ASU}{Arizona State University, Tempe, AZ  85287}
\newcommand{\BenGurion}{Department of Physics, Ben-Gurion University, Be'er Sheva 84105, Israel}
\newcommand{\BNL}{Brookhaven National Laboratory, Upton, NY 11973}
\newcommand{\Brown}{Brown University, Providence, RI 02912}
\newcommand{\Bub}{Boston University, Boston, MA 02215}
\newcommand{\BU}{Boston University, Boston, MA 02215}
\newcommand{\Buffalo}{Department of Physics, University at Buffalo, SUNY Buffalo, NY 14260 USA}
\newcommand{\Caltech}{California Institute of Technology, Pasadena, CA 91125}
\newcommand{\Cardiff}{School of Physics and Astronomy, Cardiff University, The Parade, Cardiff, CF24 3AA, UK}
\newcommand{\Carleton}{Carleton University, K1S 5B6 Ottawa, Canada}
\newcommand{\Carnegie}{The Observatories of the Carnegie Institution for Science, 813 Santa Barbara St., Pasadena, CA 91101, USA}
\newcommand{\Cavendish}{Astrophysics Group, Cavendish Laboratory, J.J.Thomson Avenue, Cambridge, CB3 0HE, UK}
\newcommand{\CCA}{Center for Computational Astrophysics, 162 5th Ave, 10010, New York, NY, USA}
\newcommand{\CPPM}{Aix Marseille Univ, CNRS/IN2P3, CPPM, Marseille, France}
\newcommand{\CEADAP}{D\'epartement d’Astrophysique, CEA Saclay DSM/Irfu, 91191 Gif-sur-Yvette, France}
\newcommand{\CERN}{CERN, Geneva, Switzerland}
\newcommand{\CfA}{Harvard-Smithsonian Center for Astrophysics, MA 02138}
\newcommand{\CFT}{Center for Theoretical Physics, Polish Academy of Sciences, al. Lotnik\'{o}w 32/46, 02-668, Warsaw, Poland}
\newcommand{\Cincinnati}{University of Cincinnati, Cincinnati, OH 45221}
\newcommand{\CITA}{Canadian Institute for Theoretical Astrophysics, University of Toronto, Toronto, ON M5S 3H8, Canada}
\newcommand{\CNRSA}{CNRS, Laboratoire d'Annecy-le-Vieux de Physique Th\'{e}orique, France}
\newcommand{\CNYang}{C.N. Yang Institute for Theoretical Physics State University of New York Stony Brook, NY 11794}
\newcommand{\CMUCosmo}{Department 
of Physics, McWilliams Center for Cosmology, Carnegie Mellon University}
\newcommand{\Columbia}{Columbia University, New York, NY 10027}
\newcommand{\Cornell}{Cornell University, Ithaca, NY 14853}
\newcommand{\CPthree}{CP3-Origins, 5230 Odense, Denmark}
\newcommand{\CWRU}{Case Western Reserve University, Cleveland, OH 44106}
\newcommand{\daa}{Department of Astronomy and Astrophysics, University of Toronto, ON, M5S3H4}
\newcommand{\damtp}{DAMTP, Centre for Mathematical Sciences, Wilberforce Road, Cambridge, UK, CB3 0WA}
\newcommand{\DESY}{DESY,  22607 Hamburg, Germany}
\newcommand{\DFI}{Departamento de F\'isica, FCFM, Universidad de Chile, Blanco Encalada 2008, Santiago, Chile}
\newcommand{\DOE}{US. Department of Energy, Germantown, MD 20874}
\newcommand{\drexel}{Drexel University, Philadelphia, PA 19104}
\newcommand{\Duke}{Duke University and Triangle Universitites Nuclear Laboratory, Durham, NC 27708}
\newcommand{\DukePhys}{Department of Physics, Duke University, Durham, NC 27708, USA}
\newcommand{\dunlap}{Dunlap Institute for Astronomy and Astrophysics, University of Toronto, ON, M5S3H4}
\newcommand{\Durham}{Department of Physics, Lower Mountjoy, South Rd, Durham DH1 3LE, United Kingdom}
\newcommand{\ED}{University of Edinburgh, EH8 9YL Edinburgh, United Kingdom}
\newcommand{\EPFL}{Institute of Physics, Laboratory of Astrophysics, Ecole Polytechnique Fédérale de Lausanne (EPFL), Observatoire de Sauverny, 1290 Versoix, Switzerland}
\newcommand{\ETH}{ETH Zurich, Institute for Particle Physics, 8093 Zurich, Switzerland}
\newcommand{\FNAL}{Fermi National Accelerator Laboratory, Batavia, IL 60510}
\newcommand{\FQAUB}{Dept. de F\' isica Qu\` antica i Astrof\' isica, Universitat de Barcelona, Mart\' i i Franqu\` es 1, E08028 Barcelona, Spain}
\newcommand{\FSU}{Florida State University, Tallahassee, FL 32306}
\newcommand{\Glasgow}{University of Glasgow, G12 8QQ Glasgow, United Kingdom}
\newcommand{\GRAPPA}{GRAPPA Institute, University of Amsterdam, Science Park 904, 1098 XH Amsterdam, The Netherlands}
\newcommand{\GSFC}{Goddard Space Flight Center, Greenbelt, MD 20771 USA}
\newcommand{\GWU}{George Washington University, Washington, DC 20052}
\newcommand{\Hampton}{Hampton University, Hampton, VA 23668}
\newcommand{\HarvardPhys}{Department of Physics, Harvard University, Cambridge, MA 02138, USA}
\newcommand{\Haverford}{Haverford College, 370 Lancaster Ave, Haverford PA, 19041, USA}
\newcommand{\Hawaii}{University of Hawaii, Honolulu, HI 96822}
\newcommand{\HKUST}{The Hong Kong University of Science and Technology, Hong Kong SAR, China}
\newcommand{\houston}{University of Houston, Houston, TX 77204}
\newcommand{\IAP}{Institut d'Astrophysique de Paris (IAP), CNRS \& Sorbonne University, Paris, France}
\newcommand{\IAS}{Institute for Advanced Study, Princeton, NJ 08540}
\newcommand{\IBS}{Institute for Basic Science (IBS), Daejeon 34051, Korea}
\newcommand{\ICC}{ICC, University of Barcelona, IEEC-UB, Mart\' i i Franqu\` es, 1, E08028 Barcelona, Spain}
\newcommand{\ICCD}{Institute for Computational Cosmology, Department of Physics, Durham University, South Road, Durham, DH1 3LE, UK}
\newcommand{\ICE}{Institute of Space Sciences (ICE, CSIC), Campus UAB, Carrer de Can Magrans, s/n, 08193 Barcelona, Spain}
\newcommand{\ICRR}{Institute for Cosmic Ray Resaerch, The University of Tokyo, 456 Higashi-Mozumi, Kamioka, Hida, Gifu 506-1205, Japan}
\newcommand{\ICTP}{International Centre for Theoretical Physics, Strada Costiera, 11, I-34151 Trieste, Italy}
\newcommand{\IFAE}{Institut de Fisica d’Altes Energies, The Barcelona Institute of Science and Technology, Campus UAB, 08193 Bellaterra (Barcelona), Spain}
\newcommand{\IFPU}{IFPU - Institute for Fundamental Physics of the Universe, Via Beirut 2, 34014 Trieste, Italy}
\newcommand{\IFT}{Instituto de Fisica Teorica UAM/CSIC, Universidad Autonoma de Madrid, 28049 Madrid, Spain}
\newcommand{\IFUNAM}{IFUNAM - Instituto de F\'{i}sica, Universidad Nacional Aut\'onoma de M\'etico, 04510 CDMX, M\'exico}
\newcommand{\IHEP}{Institute of High Energy Physics, Austrian Academy of Sciences, 1050 Vienna, Austria}
\newcommand{\Imperial}{Theoretical Physics, Blackett Laboratory, Imperial College, London, SW7 2AZ, U.K.}
\newcommand{\Indiana}{Indiana University, Bloomington, IN 47405}
\newcommand{\INAFOATs}{INAF - Osservatorio Astronomico di Trieste, Via G.B. Tiepolo 11, 34143 Trieste, Italy}
\newcommand{\INAFOAS}{INAF - Osservatorio di Astrofisica e Scienza dello Spazio di Bologna, via Piero Gobetti 93/3, I-40129 Bologna, Italy}
\newcommand{\INFNCag}{Istituto Nazionale di Fisica Nucleare, Sezione di Cagliari,  09126 Cagliari, Italy}
\newcommand{\INFNCat}{Istituto Nazionale di Fisica Nucleare, Sezione di Catania, 95125 Catania, Italy}
\newcommand{\INFNG}{Istituto Nazionale di Fisica Nucleare, Sezione di Genova, 16146 Genova, Italy}
\newcommand{\INFN}{INFN – National Institute for Nuclear Physics, Via Valerio 2, I-34127 Trieste, Italy}
\newcommand{\INFNFE}{Istituto Nazionale di Fisica Nucleare, Sezione di Ferrara, 40122, Italy }
\newcommand{\INFNLNF}{Istituto Nazionale di Fisica Nucleare, Laboratori Nazionali di Frascati, 00044 Frascati, Italy}
\newcommand{\INFNLNS}{Istituto Nazionale di Fisica Nucleare, Laboratori Nazionali del Sud, 95125 Catania, Italy}
\newcommand{\INFNN}{Istituto Nazionale di Fisica Nucleare, Sezione di Napoli, 80125 Napoli, Italy }
\newcommand{\INFNRM}{Istituto Nazionale di Fisica Nucleare, Sezione di Roma, 00185 Roma, Italy}
\newcommand{\INFNT}{Istituto Nazionale di Fisica Nucleare, Sezione di Torino, 10125, Italy }
\newcommand{\ioa}{Institute of Astronomy, University of Cambridge,Cambridge CB3 0HA, UK}
\newcommand{\IPP}{Institute for Particle Physics, BC V8W 3P6 Victoria, Canada}
\newcommand{\IPMU}{Kavli Insitute for the Physics and Mathematics of the Universe (WPI), University of Tokyo, 277-8583 Kashiwa , Japan}
\newcommand{\IPNL}{Universit\'e de Lyon, F-69622, Lyon, France; Universit\'e de Lyon 1, Villeurbanne; CNRS/IN2P3, Institut de Physique Nucl\'eaire de Lyon}
\newcommand{\IRFU}{IRFU, CEA, Universit\'e Paris-Saclay, F-91191 Gif-sur-Yvette, France}
\newcommand{\ITFA}{Institute for Theoretical Physics, University of Amsterdam, Science Park 904, 1098 XH Amsterdam, The Netherlands}
\newcommand{\IUCAA}{The Inter-University Centre for Astronomy and Astrophysics, Pune, 411007, India}
\newcommand{\Jerusalem}{Hebrew University of Jerusalem, 91904 Jerusalem, Israel}
\newcommand{\JHU}{Johns Hopkins University, Baltimore, MD 21218}
\newcommand{\JLAB}{Thomas Jefferson National Laboratory, Newport News, VA 23606}
\newcommand{\JPL}{Jet Propulsion Laboratory, California Institute of Technology, Pasadena, CA, USA}
\newcommand{\KASSI}{Korea Astronomy and Space Science Institute, Daejeon 34055, Korea}
\newcommand{\kavli}{Kavli Institute for Cosmology, Cambridge, UK, CB3 0HA}
\newcommand{\KIAS}{School of Physics, Korea Institute for Advanced Study, 85 Hoegiro, Dongdaemun-gu, Seoul 130-722, Korea}
\newcommand{\KICP}{Kavli Institute for Cosmological Physics, Chicago, IL 60637}
\newcommand{\KIPAC}{Kavli Institute for Particle Astrophysics and Cosmology, Stanford 94305}
\newcommand{\KINGS}{King's College London, WC2R 2LS London, United Kingdom}
\newcommand{\Kobe}{Kobe University, 657-8501 Kobe, Japan}
\newcommand{\KPH}{Johannes Gutenberg University, 55128 Mainz, Germany}
\newcommand{\KPMU}{University of Tokyo, 277-8583  Kashiwa , Japan}
\newcommand{\KSU}{Kansas State University, Manhattan, KS 66506}
\newcommand{\Lafayette}{Lafayette College, Easton, PA 18042}
\newcommand{\LANL}{Los Alamos National Laboratory, Los Alamos, NM 87545}
\newcommand{\LBL}{Lawrence Berkeley National Laboratory, Berkeley, CA 94720}
\newcommand{\Leiden}{Lorentz Institute, Leiden University, Niels Bohrweg 2,Leiden, NL 2333 CA, The Netherlands}
\newcommand{\Liverpool}{University of Liverpool,  L69 7ZE Liverpool , United Kingdom}
\newcommand{\LLNL}{Lawrence Livermore National Laboratory, Livermore, CA, 94550}
\newcommand{\LPC}{Universit\'e Clermont Auvergne, CNRS/IN2P3, Laboratoire de Physique de Clermont, F-63000 Clermont-Ferrand, France}
\newcommand{\LPNHE}{Sorbonne Universit\'e, Universit\'e Paris Diderot, CNRS/IN2P3, Laboratoire de Physique Nucl\'eaire et de Hautes Energies, LPNHE, 4 Place Jussieu, F-75252 Paris, France}
\newcommand{\McGill}{McGill University, Montreal, QC H3A 2T8, Canada}
\newcommand{\Melbourne}{School of Physics, The University of Melbourne, Parkville, VIC 3010, Australia}
\newcommand{\Mines}{Colorado School of Mines, Golden, CO 80401}
\newcommand{\MIT}{Massachusetts Institute of Technology, Cambridge, MA 02139}
\newcommand{\MPE}{Max-Planck-Institut f\"{u}r extraterrestrische Physik (MPE), Giessenbachstrasse 1, D-85748 Garching bei M\"unchen, Germany}
\newcommand{\MPIA}{Max-Planck-Institut f\"{u}r Astrophysik, Karl-Schwarzschild-Str. 1, 85741 Garching, Germany}
\newcommand{\MPP}{Max-Planck-Institut f\"{u}r Physik (Werner-Heisenberg-Institut), F\"ohringer Ring 6, D-80805 M\"unchen, Germany}
\newcommand{\LUPM}{Laboratoire Univers et Particules de Montpellier, Univ. Montpellier and CNRS, 34090 Montpellier, France}
\newcommand{\NAOC}{National Astronomical Observatories, Chinese Academy of Sciences, PR China}
\newcommand{\NCBJ}{National Center for Nuclear Research, Ul.Pasteura 7,Warsaw, Poland}
\newcommand{\NCU}{National Central University, Taoyuan City 32001, Taiwan (R.O.C.)}
\newcommand{\NCSU}{Physics Department, North Carolina State Universitym, 2401 Stinson Dr, Raleigh, NC 27607}
\newcommand{\ND}{University of Notre Dame,vNotre Dame, IN 46556}
\newcommand{\NIU}{Northern Illinois University, DeKalb, Illinois 60115}
\newcommand{\NMSU}{New Mexico State University, Las Cruces, NM 88003}
\newcommand{\NOAO}{National Optical Astronomy Observatory, 950 N. Cherry Ave., Tucson, AZ 85719 USA}
\newcommand{\Northwestern}{Northwestern University, Evanston, IL 60201}
\newcommand{\Nottingham}{University of Nottingham, NG7 2RD Nottingham, United Kingdom}
\newcommand{\NWU}{Northwestern University, Evanston, IL 60208}
\newcommand{\NYU}{New York University, New York, NY 10003}
\newcommand{\OK}{ University of Oklahoma, Norman, OK 73019}
\newcommand{\ORNL}{Oak Ridge National Laboratory, Oak Ridge, TN 37831}
\newcommand{\OSU}{The Ohio State University, Columbus, OH 43212}
\newcommand{\OU}{Department of Physics and Astronomy, Ohio University, Clippinger Labs, Athens, OH 45701, USA}
\newcommand{\OskarKlein}{Oskar Klein Centre for Cosmoparticle Physics, Stockholm University, AlbaNova, Stockholm SE-106 91, Sweden}
\newcommand{\Oxford}{The University of Oxford, Oxford OX1 3RH, UK}
\newcommand{\Oxy}{Occidental College, Los Angeles, CA 90041}
\newcommand{\ParisSud}{Universit\'{e} Paris-Sud, LAL, UMR 8607, F-91898 Orsay Cedex, France \& CNRS/IN2P3, F-91405 Orsay, France}
\newcommand{\PI}{Perimeter Institute, Waterloo, Ontario N2L 2Y5, Canada}
\newcommand{\Pitt}{University of Pittsburgh and PITT PACC, Pittsburgh, PA 15260}
\newcommand{\PNNL}{Pacific Northwest National Laboratory ,Richland, WA 99352}
\newcommand{\PNPI}{Petersburg Nuclear Physics Institute, 188300 Gatchina, Russia}
\newcommand{\Port}{Institute of Cosmology \& Gravitation, University of Portsmouth, Dennis Sciama Building, Burnaby Road, Portsmouth PO1 3FX, UK}
\newcommand{\Princeton}{Princeton University, Princeton, NJ 08544}
\newcommand{\PSU}{The Pennsylvania State University, University Park, PA 16802}
\newcommand{\Purdue}{Purdue University, West Lafayette, IN 47907}
\newcommand{\PW}{Participation Worldscope, Sedona, Arizona and Hong Kong, SAR PRC}
\newcommand{\Queens}{Queen's University , K7L 3N6 Kingston, Canada}
\newcommand{\Queensland}{The University of Queensland, School of Mathematics and Physics, QLD 4072, Australia}
\newcommand{\QMUL}{Queen Mary University of London, Mile End Road, London E1 4NS, United Kingdom}
\newcommand{\RAL}{Radio Astronomy Laboratory, University of California Berkeley, Berkeley, CA 94720, USA}
\newcommand{\Rice}{Department of Physics \& Astronomy, Rice University, Houston, Texas 77005, USA}
\newcommand{\RIT}{Rochester Institute of Technology}
\newcommand{\RomaS}{Dipartimento di Fisica, Universit\`{a} La Sapienza, P. le A. Moro 2, Roma, Italy}
\newcommand{\RUG}{Kapteyn Astronomical Institute, University of Groningen, P.O. Box 800, 9700 AV Groningen, The Netherlands}
\newcommand{\Rutgers}{Department of Physics and Astronomy, Rutgers, the State University of New Jersey, 136 Frelinghuysen Road, Piscataway, NJ 08854, USA}
\newcommand{\Sanford}{Sanford Underground Research Facility, Lead, SD 57754}
\newcommand{\Sassari}{Universit\`a di Sassari, 07100 Sassari,  Italy}
\newcommand{\SCIPP}{University of California at Santa Cruz, Santa Cruz, CA 95064}
\newcommand{\Sejong}{Department of Physics and Astronomy, Sejong University, Seoul, 143-747, Korea}
\newcommand{\Sheffield}{University of Sheffield, S3 7RH Sheffield, United Kingdom}
\newcommand{\SHAO}{Shanghai Astronomical Observatory (SHAO), Nandan Road 80, Shanghai 200030, China}
\newcommand{\Siena}{Siena College, 515 Loudon Road, Loudonville, NY 12211, USA}
\newcommand{\SISSA}{SISSA - International School for Advanced Studies, Via Bonomea 265, 34136 Trieste, Italy}
\newcommand{\SimonFraser}{Department of Physics, Simon Fraser University, Burnaby, British Columbia, Canada V5A 1S6}
\newcommand{\SLAC}{SLAC National Accelerator Laboratory, Menlo Park, CA 94025}
\newcommand{\SMU}{Southern Methodist University, Dallas, TX 75275}
\newcommand{\SNOLAB}{SNOLAB, Lively, ON P3Y 1N2, Canada}
\newcommand{\SoCal}{University of Southern California, CA 90089 }
\newcommand{\Stanford}{Stanford University, Stanford, CA 94305}
\newcommand{\StonyBrook}{Stony Brook University, Stony Brook, NY 11794}
\newcommand{\STSCI}{Space Telescope Science Institute, Baltimore, MD 21218}
\newcommand{\SUNYA}{University at Albany SUNY, Albany, NY 12222}
\newcommand{\SussexAstronomy}{Astronomy Centre, School of Mathematical and Physical Sciences, University of Sussex, Brighton BN1 9QH, United Kingdom}
\newcommand{\Syracuse}{Syracuse University, Syracuse, NY 13244}
\newcommand{\Tamu}{Texas AandM University, College Station, TX 77843 }
\newcommand{\Techsource}{Techsource Incorporated, Los Alamos, NM 87544}
\newcommand{\TelAviv}{Tel-Aviv University,  69978 Tel-Aviv, Israel}
\newcommand{\Temple}{Temple University, Philadelphia, PA 19122}
\newcommand{\TIFR}{Tata Institute of Fundamental Research, Homi Bhabha Road, Mumbai 400005 India}
\newcommand{\Tsinghua}{Department of Physics and Tsinghua Center for Astrophysics, Tsinghua University, Beijing 100084, P R China}
\newcommand{\TUM}{Technical University of Munich,  80333 Munich, Germany}
\newcommand{\UA}{University of Alabama, Tuscaloosa, AL 35487}
\newcommand{\UAS}{Department of Astronomy/Steward Observatory, University of Arizona, Tucson, AZ  85721}
\newcommand{\UAM}{Universidad Aut\'onoma de Madrid, 28049, Madrid, Spain}
\newcommand{\UBC}{University of British Columbia, Vancouver, BC V6T 1Z1, Canada}
\newcommand{\UCB}{Department of Astronomy, University of California Berkeley, Berkeley, CA 94720, USA}
\newcommand{\UCBP}{Department of Physics, University of California Berkeley, Berkeley, CA 94720, USA}
\newcommand{\UCBSSL}{Space Sciences Laboratory, University of California Berkeley, Berkeley, CA 94720, USA}
\newcommand{\UCD}{University of California at Davis, Davis, CA 95616}
\newcommand{\UChicago}{University of Chicago, Chicago, IL 60637}
\newcommand{\UCI}{University of California, Irvine, CA 92697}
\newcommand{\UCLA}{University of California at Los Angeles, Los Angeles,  CA 90095}
\newcommand{\UCL}{University College London, WC1E 6BT London, United Kingdom}
\newcommand{\UCR}{University of California at Riverside, Riverside, CA 92521}
\newcommand{\UCSB}{University of California at Santa Barbara, Santa Barbara, CA 93106}
\newcommand{\UCSC}{University of California at Santa Cruz, Santa Cruz, CA 95064}
\newcommand{\UCSD}{University of California San Diego, La Jolla, CA 92093}
\newcommand{\UFL}{University of Florida, Gainesville, FL 32611}
\newcommand{\UFN}{Universit\`a Federico II di Napoli, 80125 Napoli, Italy}
\newcommand{\UGTO}{Divisi\'on de Ciencias e Ingenier\'ias, Universidad de Guanajuato, Le\'on 37150, M\'exico}
\newcommand{\UKY}{University of Kentucky, Lexington, KY 40506}
\newcommand{\UMD}{University of Maryland, College Park, MD 20742
\newcommand{\UMiami}{University of Miami, Coral Gables, FL 33124}}
\newcommand{\UMich}{University of Michigan, Ann Arbor, MI 48109}
\newcommand{\UMN}{University of Minnesota, Minneapolis, MN 55455}
\newcommand{\UnB}{Instituto de F\'{i}sica, Universidade de Bras\'{i}lia, 70919-970, Bras\'{i}lia, DF, Brazil}
\newcommand{\UNC}{University of North Carolina at Chapel Hill, Chapel Hill, NC 27599}
\newcommand{\UNH}{University of New Hampshire, Durham, NH 03824}
\newcommand{\UNIMI}{Dipartimento di Fisica ``Aldo Pontremoli'', Universit\`a{} degli Studi di Milano, via Celoria 16, 20133 Milano, Italy}
\newcommand{\UNIPD}{Dipartimento di Fisica e Astronomia ``G. Galilei'',Universit\`a degli Studi di Padova, via Marzolo 8, I-35131, Padova, Italy}
\newcommand{\UNM}{University of New Mexico, Albuquerque, NM 87131}
\newcommand{\UNV}{University of Nevada, Reno, NV 89557}
\newcommand{\UoM}{Jodrell Bank Center for Astrophysics, School of Physics and Astronomy, University of Manchester, Oxford Road, Manchester, M13 9PL, UK}
\newcommand{\UPenn}{Department of Physics and Astronomy, University of Pennsylvania, Philadelphia, Pennsylvania 19104, USA}
\newcommand{\UR}{Department of Physics and Astronomy, University of Rochester, 500 Joseph C. Wilson Boulevard, Rochester, NY 14627, USA}
\newcommand{\UrbanaC}{Department of Physics, University of Illinois at Urbana-Champaign, Urbana, Illinois 61801, USA}
\newcommand{\USC}{The University of South Carolina, Columbia, SC 29208}
\newcommand{\USD}{The University of South Dakota, Vermillion, SD 57069}
\newcommand{\UTD}{University of Texas at Dallas, Texas 75080}
\newcommand{\Utenn}{The University of Tennessee, Knoxville, TN 37996}
\newcommand{\Utah}{University of Utah, Department of Physics and Astronomy, 115 S 1400 E, Salt Lake City, UT 84112, USA}
\newcommand{\UVA}{University of Virginia, Charlottesville, VA 22903}
\newcommand{\Uvic}{University of Victoria, BC V8P 5C2 Victoria, Canada}
\newcommand{\UWaterloo}{Department of Physics and Astronomy, University of Waterloo, 200 University Ave W, Waterloo, ON N2L 3G1, Canada}
\newcommand{\UWMadison}{Department of Physics, University of Wisconsin - Madison, Madison, WI 53706}
\newcommand{\UW}{University of Washington, Seattle 98195}
\newcommand{\UWC}{Department of Physics \& Astronomy, University of the Western Cape, Cape Town 7535, South Africa}
\newcommand{\Vanderbilt}{Physics \& Astronomy Department, Vanderbilt University, PMB 401807, 2301 Vanderbilt Place, Nashville, TN 37235}
\newcommand{\VSI}{Van Swinderen Institute for Particle Physics and Gravity, University of Groningen, Nijenborgh 4, 9747~AG~Groningen, The~Netherlands}
\newcommand{\VT}{Virginia Tech, Blacksburg, VA 24061}
\newcommand{\VUU}{Virginia Union University, Richmond, Virginia, 23220}
\newcommand{\WCA}{Centre for Astrophysics, University of Waterloo, Waterloo, Ontario N2L 3G1, Canada}
\newcommand{\Weizmann}{Weizmann Institute of Science, 76100 Rehovot, Israel}
\newcommand{\Wellesley}{Wellesley College, Wellesley, MA 02481}
\newcommand{\wiscIce}{University of Wisconsin, Madison, WI 53706}
\newcommand{\WM}{College of William and Mary, Newport News, VA 23606}
\newcommand{\WUSL}{Washington University in St Louis, St. Louis, MO 63130}
\newcommand{\WVU}{CSEE, West Virginia University, Morgantown, WV 26505, USA}
\newcommand{\WVUGWAC}{Center for Gravitational Waves and Cosmology, West Virginia University, Morgantown, WV 26505, USA}
\newcommand{\Wyoming}{Department of Physics and Astronomy, University of Wyoming, Laramie, WY 82071, USA}
\newcommand{\Yale}{Department of Physics, Yale University, New Haven, CT 06520}
\newcommand{\YorkU}{Department of Physics and Astronomy, York University, Toronto, Ontario M3J 1P3, Canada}
\newcommand{\Amsterdam}{Department of Physics, Science Park, University of Amsterdam - the Netherlands}
\newcommand{\Stockholm}{The Oskar Klein Centre for Cosmoparticle Physics,
Department of Physics, Stockholm University, SE-106 91 Stockholm, Sweden}
\newcommand{\Stonybrook}{Stonybrook}
\newcommand{\lancaster}{Consortium for Fundamental Physics, Physics Department, Lancaster University, Lancaster LA1 4YB, UK}
\newcommand{\Unige}{Department of Theoretical Physics and Center for Astroparticle
Physics, University of Geneva, 24 quai E. Ansermet, CH-1211 Geneva 4, Switzerland}
\newcommand{\IAPdeux}{Institut Lagrange de Paris, Sorbonne Universit ́es, 98 bis Boulevard Arago, 75014 Paris, France}
\newcommand{\Sussex}{Department of Physics and Astronomy, University of Sussex, Brighton BN1 9QH, United Kingdom}
\newcommand{\Chile}{Grupo de Cosmolog ́ıa y Astrof ́ısica Teo ́rica, Departamento de F ́ısica,
FCFM, Universidad de Chile, Blanco Encalada 2008, Santiago, Chile}
\newcommand{\SEJONG}{Department of Physics and Astronomy, Sejong University, Seoul, 143-747}
\newcommand{\NBI}{The Niels Bohr Institute and Discovery Center, Blegdamsvej 17, DK-2100 Copenhagen, Denmark}
{\raggedright
\huge
Astro2020 Science White Paper \linebreak

Primordial Non-Gaussianity \linebreak
\normalsize

\noindent \textbf{Thematic Areas:}
Cosmology and Fundamental Physics \linebreak

\textbf{Principal Author:}

Name: P. Daniel Meerburg
 \linebreak						
Institution: University of Cambridge  
 \linebreak
Email: pdm46@cam.ca.uk
 \linebreak

\textbf{Abstract:} Our current understanding of the Universe is established through the pristine measurements of structure in the cosmic microwave background (CMB) and the distribution and shapes of galaxies tracing the large scale structure (LSS) of the Universe. One key ingredient that underlies cosmological observables is that the field that sources the observed structure is assumed to be initially Gaussian with high precision.
Nevertheless, a minimal deviation from Gaussianity is  perhaps the most robust theoretical prediction of models that explain the observed Universe; it is necessarily present even in the simplest scenarios. In addition, most inflationary models produce far higher levels of non-Gaussianity.
Since 
non-Gaussianity 
directly probes the dynamics in the early Universe, a detection would present a monumental discovery in cosmology, providing clues about physics at energy scales as high as the GUT scale. 

This white paper aims to motivate a continued search to obtain evidence for deviations from Gaussianity in the primordial Universe. 
Since the previous decadal, important advances have been made, both theoretically and observationally, which have further established the importance of deviations from Gaussianity in cosmology. Foremost, {\it primordial} non-Gaussianities are now very tightly constrained by the CMB. 
Second, models motivated by stringy physics suggest detectable signatures of primordial non-Gaussianities with a unique shape which has not been considered in previous searches. 
Third, improving constraints using LSS requires a better understanding how to disentangle non-Gaussianities sourced at late times from those sourced by the physics in the early Universe. The development of the Effective Field Theory of Large Scale Structure and a number of proposed methods to `reconstruct' the initial conditions have contributed significantly to that effort. Lastly, a new technique that utilizes multiple tracers to cancel sample variance in the biased power spectrum, promises constraints on local non-Gaussianities beyond those achievable with higher $n$-point functions in both the CMB and LSS within the coming decade.

\pagebreak

\textbf{Authors\hskip1pt/\hskip1ptEndorsers\footnoteB[1]{Names in bold indicate significant contribution. }:}



{\bf Daniel Green}\footnote{\UCSD\label{UCSD}},  Muntazir Abidi\footnote{\damtp\label{damtp}}, Mustafa A. Amin\footnote{\Rice}, Peter Adshead\footnote{\UrbanaC\label{UC}}, Zeeshan Ahmed\footnote{\SLAC\label{SLAC}}, David Alonso\footnote{\Oxford}, Behzad Ansarinejad\footnote{\Durham\label{Durham}}, 
 Robert Armstrong\footnote{\LLNL}, 
 Santiago Avila\footnote{\UAM\label{UAM}},
 Carlo Baccigalupi\footnote{\SISSA}\textsuperscript{,}\footnote{\IFPU}\textsuperscript{,}\footnote{\INFN}, Tobias Baldauf\textsuperscript{\ref{damtp}}, Mario Ballardini\footnote{\UWC}, Kevin Bandura \footnote{\WVU}\textsuperscript{,}\footnote{\WVUGWAC},
 Nicola Bartolo\footnote{\UNIPD\label{UNIPD}}, Nicholas Battaglia\footnote{\Cornell\label{Cornell}}, {\bf Daniel Baumann}\footnote{\Amsterdam\label{Amsterdam}}, Chetan Bavdhankar\footnote{\NCBJ}, Jos\'{e} Luis Bernal\footnote{\ICC}\textsuperscript{,}\footnote{\FQAUB}, Florian Beutler\footnote{\Port\label{port}},  Matteo Biagetti\textsuperscript{\ref{Amsterdam}}, Colin Bischoff\footnote{\Cincinnati\label{Cincinnati}},
 Jonathan Blazek\footnote{\EPFL\label{EPFL}}\textsuperscript{,}\footnote{\OSU\label{OSU}}, J. Richard Bond\footnote{\CITA\label{CITA}}, 
 Julian Borrill\footnote{\LBL\label{LBL}}, Fran\c{c}ois R. Bouchet\footnote{\IAP\label{IAP}},
 Philip Bull\footnote{\QMUL}, Cliff Burgess\footnote{\PI\label{PI}}, Christian Byrnes\footnote{\SussexAstronomy}, Erminia Calabrese\footnote{\Cardiff}, John E.\ Carlstrom\footnote{\UChicago\label{UChicago}}\textsuperscript{,}\footnote{\KICP\label{KICP}}\textsuperscript{,}\footnote{\ANLHEP\label{ANLHEP}}, 
 Emanuele Castorina\footnote{\UCBP}, Anthony Challinor\footnote{\ioa\label{ioa}}\textsuperscript{,}\textsuperscript{\ref{damtp}}\textsuperscript{,}\footnote{\kavli\label{kavli}}, Tzu-Ching Chang\footnote{\JPL\label{JPL}}, 
Jon\'{a}s Chaves-Montero\textsuperscript{\ref{ANLHEP}}, Xingang Chen\footnote{\CfA\label{Cfa}},  Christophe Y\`eche\footnote{\IRFU}, 
 Asantha Cooray\footnote{\UCI}, William Coulton\textsuperscript{\ref{kavli}}\textsuperscript{,}\textsuperscript{\ref{ioa}}, Thomas Crawford\textsuperscript{\ref{UChicago}}\textsuperscript{,}\textsuperscript{\ref{KICP}}, {\bf Elisa Chisari}\footnote{\Oxford}, Francis-Yan Cyr-Racine\footnote{\HarvardPhys\label{Harvard}}\textsuperscript{,}\footnote{\UNM}, Guido D'Amico \footnote{\Stanford\label{stanford}}, Paolo de Bernardis \footnote{\RomaS\label{RomaS}}\textsuperscript{,}\footnote{\INFNRM\label{INFNRM}}, Axel de la Macorra\footnote{\IFUNAM\label{IFUNAM}}, Olivier Dor\'e\textsuperscript{\ref{JPL}}, Adri Duivenvoorden\footnote{\OskarKlein\label{Stockholm}}, Joanna Dunkley\footnote{\Princeton\label{Princeton}}, {\bf Cora Dvorkin}\textsuperscript{\ref{Harvard}}, Alexander Eggemeier \textsuperscript{\ref{Durham}}, Stephanie Escoffier\footnote{\CPPM}, Tom Essinger-Hileman\footnote{\GSFC\label{GSFC}}, Matteo Fasiello\textsuperscript{\ref{port}}, Simone Ferraro\textsuperscript{\ref{LBL}}, {\bf Raphael Flauger}\textsuperscript{\ref{UCSD}},  Andreu Font-Ribera\footnote{\UCL\label{UCL}}, {\bf Simon Foreman}\footnote{\CITA}, Oliver Friedrich\textsuperscript{\ref{kavli}}, Juan Garc\'ia-Bellido\textsuperscript{\ref{UAM}}, Martina Gerbino\footnote{\UMich\label{Umich}}, 
 Vera Gluscevic\footnote{\UFL},
 {\bf Garrett Goon}\textsuperscript{\ref{damtp}}, Krzysztof M. G\'orski\textsuperscript{\ref{JPL}}, Jon E. Gudmundsson\textsuperscript{\ref{Stockholm}}, Nikhel Gupta \footnote{\Melbourne}, Shaul Hanany\footnote{\UMN\label{UMN}}, 
Will Handley,\textsuperscript{\ref{kavli}}\textsuperscript{,}\footnote{\Cavendish}, 
 Adam J. Hawken\footnote{\CPPM}, 
 J.~Colin~Hill\footnote{{\IAS\label{IAS}}}\textsuperscript{,}\footnote{\CCA}, 
 Christopher M. Hirata\textsuperscript{\ref{OSU}}, Ren\'ee Hlo\v{z}ek\footnote{\dunlap\label{dunlap}}\textsuperscript{,}\footnote{\daa}, Gilbert Holder\textsuperscript{\ref{UC}}, Dragan Huterer\textsuperscript{\ref{Umich}}, 
 Marc Kamionkowski\footnote{\JHU}, Kirit S. Karkare\textsuperscript{\ref{UChicago}}\textsuperscript{,}\textsuperscript{\ref{KICP}}, Ryan E. Keeley\footnote{\KASSI\label{KASSI}}, William Kinney\footnote{\Buffalo}, Theodore Kisner\textsuperscript{\ref{LBL}}, Jean-Paul Kneib\textsuperscript{\ref{EPFL}}, 
Lloyd Knox\footnote{\UCD}, Savvas M. Koushiappas\footnote{\Brown}, 
 Ely D.~Kovetz\footnote{\BenGurion}, Kazuya Koyama\textsuperscript{\ref{port}},  Benjamin L'Huillier\textsuperscript{\ref{KASSI}}, 
 Ofer Lahav\textsuperscript{\ref{UCL}}, Massimiliano Lattanzi\footnote{\INFNFE}, 
{\bf Hayden Lee}\textsuperscript{\ref{Harvard}}, Michele Liguori\textsuperscript{\ref{UNIPD}}, 
 {\bf Marilena Loverde}\footnote{\CNYang}, {Mathew Madhavacheril}\textsuperscript{\ref{Princeton}}, Juan Maldacena\textsuperscript{\ref{IAS}}, {\bf M.C. David Marsh}\footnote{\Stockholm}, Kiyoshi Masui\footnote{\MIT}, Sabino Matarrese\footnote{\UNIPD}, Liam McAllister\textsuperscript{\ref{Cornell}}, Jeff McMahon\textsuperscript{\ref{Umich}}, Matthew McQuinn\footnote{\UW}, Joel Meyers\footnote{\SMU}, Mehrdad Mirbabayi\footnote{\ICTP}, {\bf Azadeh Moradinezhad Dizgah}\textsuperscript{\ref{Harvard},}\footnote{\Unige}, Pavel Motloch\textsuperscript{\ref{CITA}}, 
 Suvodip Mukherjee\textsuperscript{\ref{IAP}}, Julian B.~Mu\~noz\textsuperscript{\ref{Harvard}}, Adam~D.~Myers\footnote{\Wyoming},  Johanna Nagy\textsuperscript{\ref{dunlap}}, 
 Pavel Naselsky\footnote{\NBI}, 
 Federico Nati\footnote{\UPenn\label{Upenn}},  Newburgh\footnote{\Yale},
 Alberto Nicolis\footnote{\Columbia}, Michael D. Niemack\textsuperscript{\ref{Cornell}}
 Gustavo Niz\footnote{\UGTO},  Andrei Nomerotski\footnote{\BNL\label{BNL}}, Lyman Page\textsuperscript{\ref{Princeton}}, {\bf Enrico Pajer}\textsuperscript{\ref{damtp}}, Hamsa Padmanabhan\textsuperscript{\ref{CITA}}\textsuperscript{,}\footnote{\ETH}, {Gonzalo A. Palma}\footnote{\DFI}, Hiranya V. Peiris\textsuperscript{\ref{UCL}}\textsuperscript{,}\textsuperscript{\ref{Stockholm}}, Will~J. Percival \footnote{\WCA}\textsuperscript{,}\footnote{\UWaterloo}\textsuperscript{,}\textsuperscript{\ref{PI}}, Francesco Piacentni\textsuperscript{\ref{RomaS}}\textsuperscript{,}\textsuperscript{\ref{INFNRM}}, 
 {\bf Guilherme L. Pimentel}\textsuperscript{\ref{Amsterdam}}, Levon Pogosian\footnote{\SimonFraser},  Chanda Prescod-Weinstein\footnote{\UNH}, Clement Pryke\textsuperscript{\ref{UMN}}, 
 Giuseppe Puglisi\textsuperscript{\ref{stanford}}\textsuperscript{,}\footnote{\KIPAC\label{KIPAC}}, 
 Benjamin Racine\textsuperscript{\ref{Cfa}}, 
 Radek Stompor\footnote{\APC}, 
 Marco Raveri\textsuperscript{\ref{KICP}}\textsuperscript{,}\textsuperscript{\ref{UChicago}}, 
 Mathieu Remazeilles\footnote{\UoM}, 
 Gra\c{c}a Rocha\textsuperscript{\ref{JPL}}, 
 Ashley J. Ross\footnote{\OSU}, Graziano Rossi\footnote{\SEJONG}, John Ruhl\footnote{\CWRU\label{CWRU}}, Misao Sasaki\footnote{\IPMU}, Emmanuel Schaan\textsuperscript{\ref{LBL}}\textsuperscript{,}\footnote{\UCBP}, Alessandro Schillaci\footnote{\Caltech\label{Caltech}}, Marcel Schmittfull\textsuperscript{\ref{IAS}}, Neelima Sehgal\footnote{\StonyBrook}, Leonardo Senatore\textsuperscript{\ref{KIPAC}},  Hee-Jong Seo\footnote{\OU}, 
 Huanyuan Shan\footnote{\SHAO}, 
 Sarah Shandera\footnote{\PSU}, Blake D.~Sherwin\textsuperscript{\ref{damtp}}\textsuperscript{,}\textsuperscript{\ref{kavli}}, 
 {\bf Eva Silverstein}\textsuperscript{\ref{stanford}}, Sara Simon\textsuperscript{\ref{Umich}}, {\bf An\v{z}e Slosar}\textsuperscript{\ref{BNL}}, Suzanne Staggs\textsuperscript{\ref{Princeton}}, Glenn Starkman\textsuperscript{\ref{CWRU}}, Albert Stebbins\footnote{\FNAL\label{FNAL}}, Aritoki Suzuki\textsuperscript{\ref{LBL}},  Eric R. Switzer\textsuperscript{\ref{GSFC}}, Peter Timbie\footnote{\UWMadison}, Andrew J. Tolley\footnote{\Imperial},  Maurizio Tomasi\footnote{\UNIMI}, 
 Matthieu Tristram\footnote{\ParisSud}, Mark Trodden\textsuperscript{\ref{Upenn}}, Yu-Dai Tsai\textsuperscript{\ref{FNAL}}, 
 Cora Uhlemann\textsuperscript{\ref{damtp}}, Caterina Umilt\`a\textsuperscript{\ref{Cincinnati}}, Alexander van Engelen\textsuperscript{\ref{CITA}}, M. Vargas-Maga\~na\textsuperscript{\ref{IFUNAM}}, Abigail Vieregg\textsuperscript{\ref{UChicago}}, 
 {\bf Benjamin Wallisch}\textsuperscript{\ref{IAS}}\textsuperscript{,}\textsuperscript{\ref{UCSD}}, David Wands\textsuperscript{\ref{port}}, 
 {\bf Benjamin Wandelt}\textsuperscript{\ref{IAP}}, Yi Wang\footnote{\HKUST}, Scott Watson\footnote{\Syracuse}, Mark Wise\textsuperscript{\ref{Caltech}}, W.~L.~K.~Wu\textsuperscript{\ref{KICP}}, 
 Zhong-Zhi Xianyu\textsuperscript{\ref{Harvard}}, Weishuang Xu\textsuperscript{\ref{Harvard}},  Siavash Yasini\footnote{\SoCal}, Sam Young\footnote{\MPIA}, Duan Yutong\footnote{\BU}, 
Matias Zaldarriaga\textsuperscript{\ref{IAS}}, Michael Zemcov\footnote{\RIT}, Gong-Bo Zhao\footnote{\NAOC}\textsuperscript{,}\textsuperscript{\ref{port}}, 
Yi Zheng\footnote{\KIAS}, Ningfeng Zhu\textsuperscript{\ref{Upenn}}

}

\vspace{0.3cm}
\pagebreak 
\noindent

\pagebreak

\noindent
\textbf{Introduction:} Increasingly precise measurements of the Cosmic Microwave Background (CMB) and the large-scale structure (LSS) have shown that initial conditions for our Universe can be described by only a handful of parameters. Since the last decadal \cite{Komatsu:2009kd}, the {\it Planck} satellite~\cite{Ade:2015ava} has confirmed that the initial seeds of structure must have been close to Gaussian. Truly Gaussian seeds are characterized only by the power spectrum, which is currently well described by just two parameters: the overall power and scale dependence of primordial fluctuations. Yet gravity puts a lower bound on non-Gaussianity, which typically lies a few orders of magnitude below current constraints \cite{Maldacena:2002vr,Cabass:2016cgp}. A plethora of proposed models and mechanisms populate this unexplored window of non-Gaussian signals. Distinguishing among these possibilities provides a strong motivation to look for signatures beyond the current two-parameter description. Besides evident theoretical motivation, which we will elaborate on below, significant advancements in observational cosmology will allow us to obtain tighter bounds on cosmological parameters. 

The scale of inflation is a most uncertain parameter and can range across a dozen orders of magnitude without contradicting current observations. If inflation takes place at the highest energies, significant efforts in trying to detect primordial gravitational waves will triumphantly determine this scale. But if inflation takes place at lower energies, {\it Primordial non-Gaussianities} will be our unique source of information as, unlike gravitational waves, their amplitude does not diminish with energy. Hence, by complementing gravitational wave searches, the study of non-Gaussianity will provide profound {\it new information about the early Universe} by directly probing inflationary dynamics and field content at energy scales far beyond those accessible through laboratory experiments. This is precisely why early Universe cosmology is considered one of the pillars of modern physics, connecting the disciplines of fundamental theory with empirical observations. We will summarize recent theoretical developments that have derived fundamentally new predictions for primordial non-Gaussianity, highlight physics that leads to interactions between the scalar and tensor sectors and identify the general mechanisms that produce detectable levels of non-Gaussianity. Although current bounds on non-Gaussianity are impressive, we will stress that there is ample opportunity for discovery, and such a discovery would instantly present one of the most important contributions to our understanding of the early Universe. We will end by identifying new avenues in observational cosmology that are most promising in improving bounds on non-Gaussianity in the next decade.  

\vspace{0.5cm}

\noindent
\textbf{Exploring the early Universe through non-Gaussian statistics:} Deviations from Gaussianity directly translate into signatures of the dynamics and the field content driving inflation \cite{Bartolo:2001cw,Maldacena:2002vr,Acquaviva:2002ud}. Although non-Gaussian correlations are small in the simplest models of single-field slow-roll (SFSR) inflation, a much larger fraction of inflationary models is expected to produce non-Gaussianities that could be detectable. 
Currently, {\it WMAP}~\cite{Hinshaw:2012aka} and {\it Planck}~\cite{Ade:2015ava} provide the most stringent limits on a wide range of non-Gaussian shapes that could be produced during inflation; however, today's measurements are not sufficiently sensitive to suggest a particular mechanism is favored by the data. 
At the same time, our understanding of inflation is continually refined, and there is an associated need to improve our understanding of the underlying dynamics directly through constraints on higher-order correlations \cite{Bartolo:2004if,Komatsu:2009kd,Chen:2010xka}. 

Deviations from Gaussianity in the initial fluctuations are most easily measured through their effect on the bispectrum, the Fourier transform of the three-point correlation function (similar to skewness in 1D). By homogeneity and isotropy, the bispectrum is a function of the norm of three momenta (here $k_a=|\vec{k}_1|$, for $a=1,2,3$), which combine to form a triangle; its {\it shape} describes triangular configurations where the bispectrum is largest. Together with the {\it amplitude} $f_{\rm NL}$ this defines a unique bispectrum\footnoteB{Similar to the power spectrum, the bispectrum could in principle inherit scale dependence which would introduce more degrees of freedom \cite{Byrnes:2010ft,Becker:2012je}.}. 
Different physical scenarios generate distinguishable shapes and we can identify associated thresholds for the amplitude that allow us to classify the physics of inflation (and alternatives).  

Generally, bispectra are most easily visualized according to the contributions in three distinct shapes; local, equilateral and folded triangles. Physically they correspond to a shape where $k_1 \ll k_2\sim k_3$ (squeezed or local), with amplitude $f_{\rm NL}^{\rm local}$, $k_1\sim k_2\sim k_3$ (equilateral) with amplitude $f_{\rm NL}^{\rm equil}$ and $k_1+k_2 \sim k_3$ (folded) with amplitude $f_{\rm NL}^{\rm folded}$. Detectable amounts of non-Gaussianity could be produced in the following scenarios:

\vspace{-0.3cm}

\begin{itemize}[leftmargin=0.4cm]
\setlength\itemsep{-.05cm}
    \item {\bf Inflaton self-interactions} Non-gaussanity can arise from non-linear dynamics during single-field inflation.
     In the most well-studied case, these interactions also cause the fluctuations to propagate with a speed slower than the speed of light. Both a detection or an exclusion of such a signature provides a unique window into the mechanism behind inflation.   

    \item {\bf Additional light fields} Light degrees of freedom are excited from the vacuum with an amplitude set by the Hubble scale. When this degree of freedom is not the inflaton, these fluctuations freeze-out and describe isocurvature (entropy) fluctuations.  These isocurvature modes may eventually convert into isocurvature perturbations, during inflation or reheating. These conversion processes induce correlations between modes that are necessarily non-Gaussian.
    
    \item {\bf Additional heavy fields}
    Heavy degrees of freedom (e.g. particles with mass on the order of the Hubble scale during inflation, or larger) are excited during inflation but are diluted quickly after horizon crossing. However, when the inflaton couples to these additional degrees of freedom, their fluctuations can still correlate the adiabatic modes producing non-Gaussianity.
    
\end{itemize}

\vspace{-0.2cm}

All bispectra that come from fluctuations of the field that drives inflation (``single-clock'' scenarios) most strongly couple momenta of similar wavelengths. The ``squeezed limit'' of these bispectra is very restricted for adiabatic modes, which are necessarily the only fluctuations in attractor single-clock models. 
A large fraction of the parameter space for scenarios involving interactions during inflation that respect the underlying shift symmetry (i.e.\ are approximately scale-invariant) is captured by equilateral~\cite{Babich:2004gb} and orthogonal shapes~\cite{Senatore:2009gt}, where the latter is  orthogonal to equilateral. 
Examples include scenarios in which inflaton fluctuations have non-trivial self-interactions~\cite{Silverstein:2003hf,ArkaniHamed:2003uz,Alishahiha:2004eh,Chen:2006nt,Cheung:2007st,Senatore:2009gt} or couplings between the inflaton and other (potentially massive) degrees of freedom~\cite{Chen:2009we,Chen:2009zp,Tolley:2009fg, Cremonini:2010ua, Achucarro:2010da,Baumann:2011nk,Barnaby:2011pe,Arkani-Hamed:2015bza}. 
Vanilla SFSR inflation necessarily produces $f_{\rm NL}^{\rm equil} < 1$~\cite{Creminelli:2003iq} and therefore {\it any detection of $f_{\rm NL}^{\rm equil} \geq 1$ would rule out a large class of models} and would imply that inflation is a strongly coupled phenomenon and/or involved more than one field~\cite{Baumann:2014cja,Alvarez:2014vva,Baumann:2015nta}.  

In single-field inflation, $f_{\rm NL}$ typically is related to a new energy scale, $M$, such that $f_{\rm NL}^{\rm equil} \propto \, (H/M)^2$~\cite{Cheung:2007st,Baumann:2011su}, with $H$ the hubble scale during inflation. At this energy scale self-interactions become strongly coupled and current limits on the bispectrum \cite{Ade:2015ava} translate into $M >\mathcal{O}(10)H$. In the presence of additional fields besides the inflaton, $f^{\rm equil}_{\rm NL}$ scales with the strength of the coupling between the inflaton and these additional fields, usually suppressed by an energy scale $\Lambda$. Current limits give $\Lambda  > \mathcal{O}(10{-}10^{5}) H$~\cite{Green:2013rd,Assassi:2013gxa}. Fixing the amplitude of scalar perturbations to its observed value, the tensor-to-scalar ratio $r \propto H^2$, and for $r > 0.01$ these constraints require some of the interactions to be {\it weaker than gravitational.}

When light degrees of freedom other than the inflaton contribute to the observed scalar fluctuations (i.e. multi-field inflation), coupling between modes of very different wavelengths is allowed. Historically, the most well-studied bispectrum is the local bispectrum, which couples short wavelength modes $k_2\sim k_3$ to long wavelength modes $k_1$. {\it A detection of this shape with an amplitude of $f_{\rm NL}^{\rm local} \sim {\cal O}(1)$ would rule out all attractor models of single-clock inflation} \cite{Creminelli:2004yq}. Non-attractor models exist that generate observable $f_{\rm NL}^{\rm local}$ \cite{Kinney:2005vj,Namjoo:2012aa,Martin:2012pe,Chen:2013aj,Huang:2013lda,Mooij:2015yka} and are under continued investigation \cite{Bravo:2017gct,Bravo:2017wyw,Finelli:2017fml,Cai:2017bxr,Passaglia:2018ixg}.

 Multi-field inflationary models can produce observably large local non-Gaussianity and provide a well-motivated framework for interpreting upcoming observations. It has long been known that substantial levels of non-Gaussianity can be generated after the end of inflation \cite{Lyth:2002my,Enqvist:2004ey, Bartolo:2003jx,Zaldarriaga:2003my,Lyth:2005qk}, and $f_{\rm NL}^{\rm local} \sim {\cal O}(1)$ is a natural outcome when the primordial perturbations are generated by a so-called ‘spectator’ field 
 \cite{Linde:2012bt,Meyers:2013gua, Elliston:2014zea,dePutter:2016trg,Torrado:2017qtr}. Generating observational levels of local non-Gaussianity \emph{during} multi-field inflation is more challenging, as can be understood from simple toy models \cite{Battefeld:2006sz}, general arguments \cite{Byrnes:2008wi, Byrnes:2008zy,Byrnes:2010em,Peterson:2010mv,Peterson:2011yt}, and explicit solutions of inflationary models with many interacting fields \cite{Frazer:2011br, McAllister:2012am, Bjorkmo:2017nzd}. Consequently, substantial multi-field contributions to the primordial curvature perturbations do not guarantee large non-Gaussianities, and {\it a detection of $f_{\rm NL}^{\rm local} \sim {\cal O}(1)$ would provide decisive insights into the origin of the primordial density perturbations.} Non-inflationary cosmologies can also produce large primordial non-Gaussianities of the local shape \cite{Lehners:2009ja}, and would be heavily constrained by improved limits on $f_{\rm NL}^{\rm local}$. Finally, we note that a detection of $f_{\rm NL}^{\rm local}$ would open the door to significant cosmic variance on all scales from coupling of fluctuations within our observed volume to any super-Hubble modes \cite{Nelson:2012sb,LoVerde:2013xka,Nurmi:2013xv,LoVerde:2013dgp}. Indeed, there would be room for a significant shift between the observed amplitude of scalar fluctuations (and so the observed tensor-to-scalar ratio~$r$) and the mean value of fluctuations on much larger scales \cite{Bonga:2015urq}. 

Additional theoretically well-motivated shapes are not captured by local, equilateral, folded and orthogonal triangles. For example, in models in which the inflaton is an axion with monodromy \cite{Silverstein:2008sg,McAllister:2008hb,Flauger:2009ab,Berg:2009tg}, bursts of particle or string production naturally lead to {\it periodic features} in the bispectrum where the frequency of the feature can be linked to the axion decay constant \cite{Flauger:2010ja,Leblond:2010yq,Flauger:2014ana}. Often these contributions will lead to counterparts in the power spectrum and are expected to be detected there first~\cite{Behbahani:2011it}, but this need not be the case~\cite{Behbahani:2012be}. Various other mechanisms could also introduce non-trivial features in the primordial bispectrum \cite{Chen:2006xjb,Chen:2008wn,Holman:2007na,Meerburg:2009ys,Meerburg:2009fi,Adshead:2011jq,Barnaby:2011qe,Meerburg:2012id,Achucarro:2012fd,DAmico:2012wal,Palma:2014hra}, providing a rich phenomenology in bispectrum space.   

The Hubble scale during inflation might have been as high as $10^{14}$ GeV, providing access to physics far beyond the reach of conventional particle colliders. At these energies, new massive particles, if they exist, are created by the rapid expansion of the 
inflationary space-time. When these particles decay, they can produce nontrivial correlations in the inflationary perturbations~\cite{Chen:2009zp, Baumann:2011nk, Assassi:2012zq, Chen:2012ge, Noumi:2012vr, Baumann:2012bc, Assassi:2013gxa, Arkani-Hamed:2015bza, Lee:2016vti, Kehagias:2017cym, Kumar:2017ecc, An:2017hlx, An:2017rwo, Baumann:2017jvh, Arkani-Hamed:2018kmz,Dimastrogiovanni:2014ina,Dimastrogiovanni:2015pla}.  The characteristic signature of these new particles is a non-analytic scaling in the squeezed limit of the bispectrum or the collapsed limit of the trispectrum (the Fourier transform of the 4-point function). For masses above the inflationary Hubble scale, the signal will oscillate 
and frequencies of these oscillations encode the {\it masses of the new particles.} 

Thus far, both theoretically and observationally, correlators involve only scalar degrees of freedom. However, in light of upcoming B-mode polarization experiments, in principle bispectra involving multiple tensors (e.g. the scalar-scalar-tensor bispectrum (SST)) can be constrained for the first time. 
Massive particles with spin generate a nontrivial angular dependence in the squeezed limit. Certain types of spinning particles---so-called partially massless (PM) particles---can lead to an enhanced signal in the SST bispectrum~\cite{Baumann:2017jvh}. This would be a characteristic signature of the inflationary de Sitter spacetime, since PM particles have no analog in flat space. Alternatively, a non-trivial signal in the SST bispectrum can arise if the kinetic terms of the spinning fields strongly break the de Sitter symmetry~\cite{Kehagias:2017cym,Bordin:2018pca,Dimastrogiovanni:2018gkl,Ozsoy:2019slf}, if position-dependent background fields break the spatial isometries  \cite{Endlich:2012pz,Cannone:2014uqa,Cannone:2015rra,Ricciardone:2016lym,Piazza:2017bsd} or, more generally, if the tensors are sourced by additional field, e.g. in gauge-flation \cite{Barnaby:2010vf,Barnaby:2011pe,Barnaby:2012xt,Maleknejad:2011jw, Adshead:2016iix}). 
Non-Gaussian signals may also arise from particles within the Standard Model~\cite{Chen:2016hrz, Chen:2016uwp, Chen:2016nrs}. For instance, if the Higgs field has a coupling to curvature, it can acquire a mass of order the Hubble scale during inflation, and naturally couple to the inflaton in pairs, contributing to non-Gaussianity. Similarly, scalar partners in supersymmetric theories would produce non-Gaussianity if they exist anywhere up to the inflationary Hubble scale~\cite{Baumann:2011nk}.

Finally, a more general question is the role of higher $n$-point functions of scalar fluctuations. For example, if the inflaton couples directly to other fields, additional particles may be produced at a mass scale up to of order the square root of the kinetic energy of the inflaton field. Axion fields in string theory introduce periodic events of this kind. The signal to noise for the resulting non-Gaussianity {\it peaks at a value of $n$ which can be greater than 3} \cite{Flauger:2016idt}. This implies a reach of observations to a higher scale than the inflationary Hubble scale. It is of interest to characterize the contribution that tails of the distribution might make to phenomenology. Early work covering aspects of this appeared in \cite{Bond:2009xx}, and several groups are investigating the problem more generally \cite{Chen:2018uul,Chen:2018brw}. The amplitude of the tails exhibits exponential sensitivity to model parameters, whose characterization requires a careful theoretical analysis. This direction, as well as additional shapes of low-point correlation functions, {\it promise to increase the physics that can be learned from the analysis of primordial non-Gaussianity.}

\vspace{0.5cm}

\noindent
\textbf{Prospects for the measurement of non-Gaussianities in the next decade:} 
{\it Planck} has provided constraints \cite{Ade:2015ava} on the most theoretically compelling shapes discussed in the previous section, improving bounds from {\it WMAP} by almost an order of magnitude \cite{Hinshaw:2012aka}. The original method to constrain the primordial bispectrum in the CMB and in LSS relied on the primordial shape being of simple factorizable form, forcing the analysis to use specifically designed templates. Leading up to {\it Planck}, new methods \cite{Fergusson:2008ra,Fergusson:2006pr,Fergusson:2010gn,Bucher:2015ura} have been developed that have opened up the space of constrained shapes dramatically. Now, almost thirty thousand different shapes have been put to the test \cite{Ade:2015ava}. Despite these improvements, bispectra that contain features have proven hard to constrain, since the frequency and phase of the features have broad theoretical priors. New methods developed better equipped to look for such bispectra \cite{Adshead:2011bw,Adshead:2011jq,Munchmeyer:2014cca,Meerburg:2015yka} have allowed the {\it Planck} collaboration to explore a significant part of this parameter space, thus far without finding significant evidence for deviations from non-Gaussianity \cite{Ade:2015ava}. In addition, since features in the power spectrum and the bispectrum generally contain correlated parameters \cite{Meerburg:2009ys,Achucarro:2010da, Flauger:2010ja,Meerburg:2015yka,Achucarro:2012fd,Palma:2014hra}, statistical methods have been developed to use constraints from both the power spectrum and the bispectrum to further constrain model space \cite{Fergusson:2014hya,Fergusson:2014tza,Meerburg:2015owa} and joint analysis of the power spectrum and bispectrum were presented in \cite{Fergusson:2014tza,Akrami:2018odb}.  

Because of its computational complexity, the search for non-Gaussianity differs from the measurement of the primordial power spectrum. Unlike the power spectrum, the bispectrum and higher order $n$-point functions are pre-calculated spectra and the cosmology is held fixed; only the shape is varied and the amplitude $f_{\rm NL}$ is determined from the data. This implies that if we have yet to determine the correct shape of the primordial bispectrum, we could very well miss the signal entirely.
On the other hand, the same richness of possible inflationary models increases the possibility of false detections due to the look-elsewhere effect. 

Various ongoing and planned CMB experiments will significantly improve polarization sensitivity and measurements down to smaller scales further constraining non-Gaussianities \cite{Abazajian:2016yjj,Ade:2018sbj,PICO}. 
It must be noted that improved sensitivity requires a careful treatment of secondary effects that are imprinted in the CMB from both extra-galactic \cite{Lewis:2011fk,Kim:2013nea,Curto:2014bna,Coulton:2017crj,Hill:2018ypf} and galactic origin \cite{Jung:2018rgf,vonHausegger:2018tjq}, which could obscure the primordial signal. The latter would benefit from using multi-frequency data \cite{Stacey:2018yqe}. Non-Gaussian contributions to the covariance can also become important \cite{Babich:2004yc,Lewis:2011fk}. Alternatively, the CMB can constrain local non-Gaussianities using spectral distortions \cite{Pajer:2012vz,Ganc:2012ae,Pajer:2013oca,Emami:2015xqa,Khatri:2015tla,Bartolo:2015fqz,Ota:2016mqd,Ravenni:2017lgw,Remazeilles:2018kqd,Cabass:2018jgj}. 

Beyond the CMB, developments in large-scale structure theory and analysis demonstrate that LSS could provide us with even better constraints than those obtained with the CMB \cite{Alvarez:2014vva,Dore:2014cca,Dodelson:2016wal}. Local non-Gaussianity uniquely produces effects on {\it both} power spectrum \cite{Dalal:2007cu,Matarrese:2008nc} and bispectrum of tracers of large-scale structure. The effect of local non-Gaussianity on LSS is relatively robust with respect to theoretical modeling because gravitational interactions cannot generate this signal. While measuring power spectra is a remarkably advanced technique in LSS analysis, from a systematic point of view, clean measurements of very large scales are particularly difficult due to imprints of our own galaxy, solar system neighbourhood and survey strategy on the observed modes. Equilateral and orthogonal shape suffer from the opposite problem; observations are likely to be cleaner, but theoretical modelling will suffer from our understanding of non-linear gravitational evolution on smaller scales. Improved perturbative understanding \cite{Angulo:2014tfa,Baldauf:2014qfa,Assassi:2015jqa,Angulo:2015eqa} of small scales will allow us to utilize more modes and improve projected constraints on the primordial correlation functions \cite{Alvarez:2014vva}. Different LSS tracers have different advantages. Galaxies from spectroscopic and photometric surveys are the most advanced tracers and will reach exquisite signal-to-noise ratios in the coming decade. Weak gravitational lensing probes dark matter directly and is theoretically easier to model. Furthermore, galaxy shapes are uniquely sensitive to anisotropy in primordial non-Gaussianity \cite{Schmidt:2015xka,Chisari:2016xki}. Neutral hydrogen traced by 21-cm allows one to go higher redshift, where the volume available is large and the universe is more linear and thus easier to model. This could significantly benefit the search for non-Gaussianities \cite{Karagiannis:2018jdt}, initially at relatively low redshifts \cite{Ansari:2018ury} and eventually throughout the entire observable universe \cite{Munoz:2015eqa}, opening up the full potential of the cosmological collider experiment \cite{Meerburg:2016zdz} when combined with low redshift probes of the LSS \cite{MoradinezhadDizgah:2017szk,MoradinezhadDizgah:2018ssw,MoradinezhadDizgah:2018pfo}. Besides neutral hydrogen, intensity mapping with other emission lines 
could further improve constraints on primordial non-Gaussianity \cite{MoradinezhadDizgah:2018zrs,MoradinezhadDizgah:2018lac}. 

Finally, recent theoretical work has shown that impressive improvements can be made when combining multiple tracers, resulting in so-called cosmic variance cancellation \cite{Seljak:2008xr}. Forecasts show \cite{Schmittfull:2017ffw,Munchmeyer:2018eey} local non-Gaussianity could be measured to levels below the theoretically motivated threshold when combining Large Synoptic Survey Telescope data \cite{2009arXiv0912.0201L} with future CMB data \cite{Ade:2018sbj}. Similar cancellation could be achieved when combining multiple measurements of the shape of galaxies in a search for anisotropic non-Gaussianity \cite{Chisari:2016xki}. 

\vspace{0.5cm}

\noindent
\textbf{Conclusion:} Though non-Gaussianity has been significantly constrained, by necessity the bounds apply only to a tiny fraction of possible non-Gaussian directions in theoretical parameter space. There is a rich
interplay between the analysis of non-Gaussianity and theoretical developments which continue to uncover novel dynamical mechanisms for inflation and its perturbations. Once data is collected, it can bear new fruit
with each additional theoretical structure that motivates novel tests. Even null results
can be very informative, illuminating the empirical boundaries in the space of well-defined theoretical
parameters. This motivates a continued effort in constraining correlation functions beyond the two-point function, which ultimately hold the only key to access physics at energy scales close to the boundary of our knowledge.

\theendnotes

\pagebreak

\bibliographystyle{unsrt}
\bibliography{whitepaper}

\end{document}